# Covid19 Vaccination Acceptance and Deprivation among US Counties

Zi Iun Lai[1], Jun Yang Ang[2,3]

[1]Department of Mechanical Engineering, Stanford University, CA, USA
[2]Department of Economics, Stanford University, CA, USA
[3]Department of Computer Science, Stanford University, CA, USA

**This report explores the central question of how socioeconomic status affects Covid19 vaccination rates in the United States, using existing open-source data. In general, a negative correlation exists between Area Deprivation Index (ADI) of a county and first dose, primary series and booster vaccination rates. Higher area deprivation correlated with polled vaccine hesitancy and lower search interest in vaccine interest, intention to vaccinate or concern about safety of vaccination. Positive correlations between ADI and certain mental health search trends were noted. No clear correlation between deprivation index and accessibility to vaccination sites were observed. In a small data sample, county level housing assistance policies and public information campaigns were noted to positively influence vaccine follow-through rates. Finally, random forest, linear regression and KNN models were explored to validate the use of the above features for vaccine acceptance prediction.**

## I. Introduction

By summer of 2021, several Covid19 vaccines, produced through Operation Warp Speed, were made available to most in the United States [1]. These vaccines are free for anyone living in the US, regardless of immigration or insurance status [2]. As of December 2023, 6.6 million hospitalizations and 1.1 million deaths due to Covid19 have been recorded [3], with the disease burden disproportionately affecting poorer communities [4]–[8]. Covid19 vaccinations likely prevented a large number of potential cases; however, over 30% of Americans have not completed primary vaccination regime [3].

Several studies have examined factors influencing Covid19 vaccination rates in the US. Through surveys of 2279 individuals conducted in April 2020, Kelly et al reported a general vaccine acceptance rate of 75% [9]. African Americans were significantly more hesitant than Hispanic or White respondents, with 47%, 80% and 75% acceptance rates respectively. Higher age was associated with higher polled willingness, with 85% acceptance among those 65 years and older, 70% among those between 25 to 34. Having existing comorbidities did not increase willingness.

In May 2020, Malik et al surveyed 672 participants and reported vaccine acceptance rate of 67% [10]. Identifying as Black, being younger, and female gender were similarly associated with lower polled acceptance. Being employed and having college education were associated with higher polled acceptance.

In a longitudinal 2020 study, Latkin et al polled 522 participants across different stages of vaccine development to identify factors contributing to vaccine distrust [11]. Rapid pace of vaccine development and approval, polarization of media and information content and social norms of their community affected trust in vaccines.

In November 2020, Shekhar analyzed 3479 responses among healthcare workers [12]. 36% of respondents surveyed would accept Covid19 vaccinations immediately, while 56% were unsure. The most common concerns were the pace of vaccine development (74%), doubts over its efficacy (69%) and safety (69%).

Loomba et al polled 3000 respondents in the US on intention to vaccinate, followed by their social media usage, sources of and attitudes towards Covid19-related content [13]. After exposure to Covid19-related content, which was either misinformation or factual, participants were again polled on their intention to vaccinate. Recent exposure to misinformation was estimated to reduce vaccine intention by up to 6.4%. The effects of misinformation varied along ethnic, gender and religious lines. Interestingly, lower income groups were less likely to decrease vaccination intentions after exposure to misinformation.

Several teams have investigated Covid19 vaccine coverage across Social Vulnerability Index (SVI) brackets. SVI is a composite metric to identify regions with higher need for assistance in event of emergencies [14]. The metric considers education and socioeconomic status, age, ethnicity, family structure, housing, and transportation arrangements of the population. Hughes et al reported that first dose vaccination coverage as of March 2021 was slightly higher among counties in the least vulnerable tertile compared to those in the most vulnerable tertile, at 15.8% and 13.9% respectively [15]. Similarly, Barry et al analyzed vaccine coverage until May 2021 across counties of different urbanicity (large metropolitan, medium-small metropolitan, fringe metropolitan and nonmetropolitan counties) and SVI quartile [16]. Fringe metropolitan and nonmetropolitan counties displayed the largest disparities in first dose coverage between the highest and lowest SVI quantiles, of approximately 13-16%. This prompted calls to explore whether initial differences in county-wide vaccine coverage across SVI arose due to access limitations or disparities in populations' responses.

On the other hand, Roghani & Panahi reported no significant trend between socioeconomic measures and vaccine coverage [17]. The researchers compared vaccination rates against state level population demographics (rates of unemployment, poverty, home ownership and education, as well as ethnicity and age). Poorer socioeconomic measures did not correspond to lower vaccination rates; in fact, unemployment was noted to be positively correlated with higher vaccination rates. However, it should be noted that the small number of low-granularity data points used in this report, as well as significant political/policy differences between states, could have obfuscated underlying trends.

Hildreth & Alcendor reviewed vaccine hesitancy among ethnic minorities in the US [18]. Researchers noted vaccine hesitancy were associated with supply chain and access inequities, hesitancy among healthcare providers, and disinformation. Among African Americans, historical prejudice was believed to have eroded trust in the healthcare system. Among Hispanic communities, fear of interacting with official institutions, misinformation and disinformation were believed to contribute to higher rates of vaccine hesitancy.

Clouston et al compared early rates of vaccination in counties of different socioeconomic statuses (SES), quantified through a composition of education, unemployment and income [19]. Analyzing vaccination data in 2021, researchers reported that an increase in standard deviation of SES corresponded with 3.8% increase in daily vaccination rates within the time period. Inequality in distribution was posited as a key reason for this disparity. Previous studies provide valuable insights into vaccine hesitancy in different demographics. This article utilizes open-source datasets to explore how regional deprivation correlates with vaccine response across US counties.

## II. Datasets Description

**Area Deprivation Index (ADI) Percentile**

The Area Deprivation Index (ADI) is a composite measure of the level of deprivation among US counties [20], [21]. ADI percentile ranges from 0-100; higher scores indicate higher deprivation. Scores are calculated using 17 indicators from the American Community Survey; these indicators include income, education, unemployment rate and category, regional income disparity, home valuation and ownership rate, rate of single parenthood, and rate of ownership of different facilities such as vehicles, telephones, plumbing. ADI has been linked to numerous causes of morbidity and mortality [22]–[25], and is used by the Health Resources & Services Administration (HRSA) to identify communities with higher needs of healthcare support. For each FIPS code, the most recent available data between the years 2018-2020 was used.

**US County Vaccination Rates**

County level vaccination rates for first dose, primary series and booster shots are made available by the Centers for Disease Control and Prevention, via Open Data Network [26]. The dataset records cumulative and new rates every epidemiological week until May 2023. The most recent record of percentage of vaccinated population per county is used, with missing or invalid entries removed.

**Covid19 Vaccination Search Insights**

Released by Google under BigQuery Public Datasets Program, the dataset consists of anonymized aggregated relative interest of Google search engine searches for 1. General interest in Covid19 vaccination, 2. Specific intentions to receive vaccinations and 3. Safety concerns regarding Covid19 vaccinations [27]. Longitudinal scores were averaged for each county.

**Covid19 Vaccine Hesitancy**

Developed by the Delphi Group at Carnegie Mellon University and Facebook, this dataset contains estimates of Covid19 vaccine hesitant percentages in different US counties [28]. Polls were conducted January-May 2021, as vaccinations were being rolled out to the public. The values corresponding to the latest dates were used.

**County level Covid19 Assistance Policies**

This dataset categorizes county level Covid19 assistance policies in 171 counties in California, Louisiana, Mississippi, New Jersey, New York, Texas, and Utah [29]. Listed counties cover >25% of US population and represent diverse ethnic/political groups. Policies analyzed spanned January-March 2021. Utility and housing support policies were categorized with '0' or '1' for absence and presence of efforts, or '9' for unknown. Public information efforts were categorized as '0', '1', '2' for no information, some information, and comprehensive information provided, or '9' for unknown. It should be noted that this dataset is very small, considering only about 5% of over 3000 counties in the US. More data and analysis are required to draw more definitive conclusions.

**Negative Twitter Sentiments on Covid19**

This dataset quantifies the extent of negative sentiments – anger, fear, sadness, disgust – relating to Covid19 in public posts on Twitter (now X) on a weekly basis [30]. Scores are normalized by the total number of Tweets. Average scores across time are used for each county to estimate sentiments.

**Covid19 Vaccination Access**

Available through BigQuery Public Datasets Program, this dataset contains locations of vaccination sites in each county, and the estimated travel duration to the site within the community it was intended to serve [31]. Data contains estimated travel time via different modes of transport, as of November 2021. However, the data seems slightly contrived, with travel time categorized into discrete buckets of 15 minutes.

**Covid19 Mental Health Symptom Searches**

The Google Covid19 Search Trends dataset contains aggregated weekly search trends for health symptoms which could relate to Covid19 [32]. Scores for each symptom are normalized to total search activity at each county. This project explores searchers related to depression, mania, insomnia and panic attacks. Data is openly available under BigQuery Public Datasets Program.

**Meal Gap Index**

Statistics on food insecurity were kindly made available through Feeding America's Map the Meal Gap (MMG) program [33], [34]. Briefly, the model utilizes data on regional meal costs and expenditures, disability, unemployment, income, homeownership rates, and ethnicity from the Current Population Survey across US counties. Data from 2020 (released in 2022) was used.

## III. Results and Visualizations

### A. Vaccination Rates against Area Deprivation Index

The following plot shows average first dose, primary series and booster vaccination rates across all counties, grouped by their ADI percentile. In general, higher regional deprivation scores correlate with lower vaccination rates, with Spearman's correlation coefficient $r_s(101)$ = -0.755, -0.795 and -0.853 for rates of first dose, primary series and booster shots respectively ($p<0.001$).

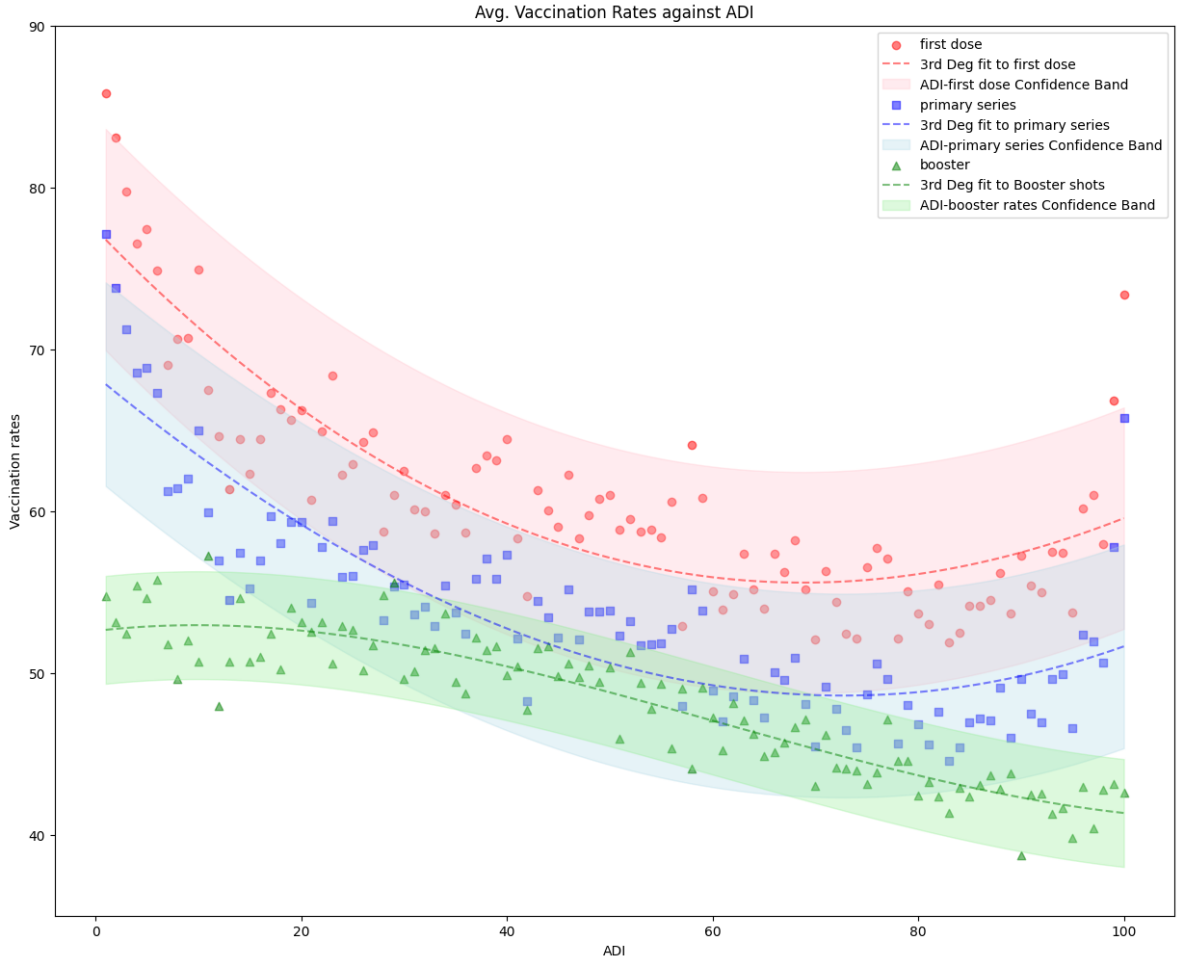

**Figure 1** – Scatter plot of average first dose rates (red), primary series completion rates (blue) and booster rates (green) of Covid19 vaccinations of counties aggregated by ADI score, with 3$^{rd}$ order best fit curves and 95% confidence intervals (CI).

We note the correlation between area deprivation and vaccination rates for booster shots exhibit a strong linear trend. Unlike first and primary vaccine doses, receiving booster shot could have been less supported by official programs, leaving it more to individual discretion. There are also outliers: the single data point at 0% ADI, indicating low deprivation, shows unusually low rates of vaccination. This sample corresponds to Falls Church City in Virginia, a predominantly White and Democratic community with a population of over 10,000. Other sources suggest first dose and primary vaccination rates for Falls Church City are approximately 93% and 86% respectively as of Dec 2022, suggesting an error in CDC dataset for this county [35]. This datapoint was neglected from analysis. On the other end of the spectrum, the 63 counties at the 99% and 100% ADI showed an unusually high rate of first dose and primary vaccinations.

B. Vaccination Site Density against ADI

Data for travel time via driving, public transit and walking in Vaccination Access dataset is highly discretized, displayed in 15-minute intervals from 15-60 mins. Nearly no difference in ADI score distribution was found when grouped by travel time duration. Number of vaccination sites correlation weakly and negatively with ADI of county ($r_s(2862) = -0.285$, $p < 0.01$). suggesting poorer counties have less vaccination sites. However, when adjusted for population size, counties with higher ADI have more vaccination sites per 10k population. This is likely related to the trend of poorer counties being rural and having smaller populations in the US [36].

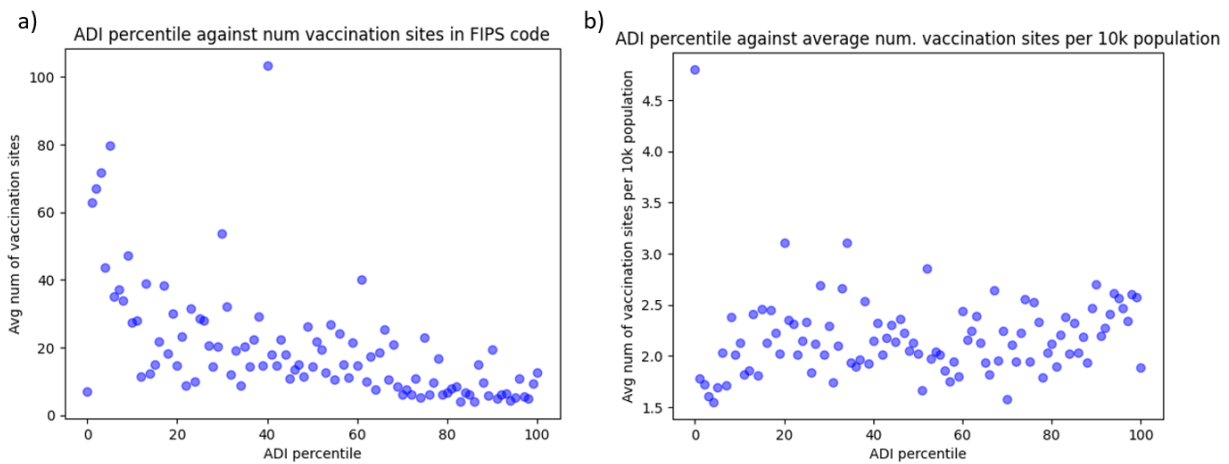

**Figure 2** – Scatter plot of a): average number of vaccination sites per county at each ADI percentile, b): average number of vaccination sites per 10,000 population for counties at each ADI percentile.

While previous studies reported that unequal access resulted in disparities in vaccination rates between rich and poor counties at the start of vaccine distribution [19], data from mid-2023 does not show strong differences in accessibility. Furthermore, positioning of vaccine sites could have been influenced by other factors, such as density and presence of critical populations (healthcare workers, elderly).

C. Vaccination Interest and Polled Intention against ADI

Vaccine interest, estimated by search terms, are aggregated according to ADI percentile below. Average vaccination score refers to relative number of Google searches on Covid19 vaccines, intention score quantifies relative searches for how to receive vaccines, and safety concern score quantifies relative number of queries about safety and side effects of Covid19 vaccines.

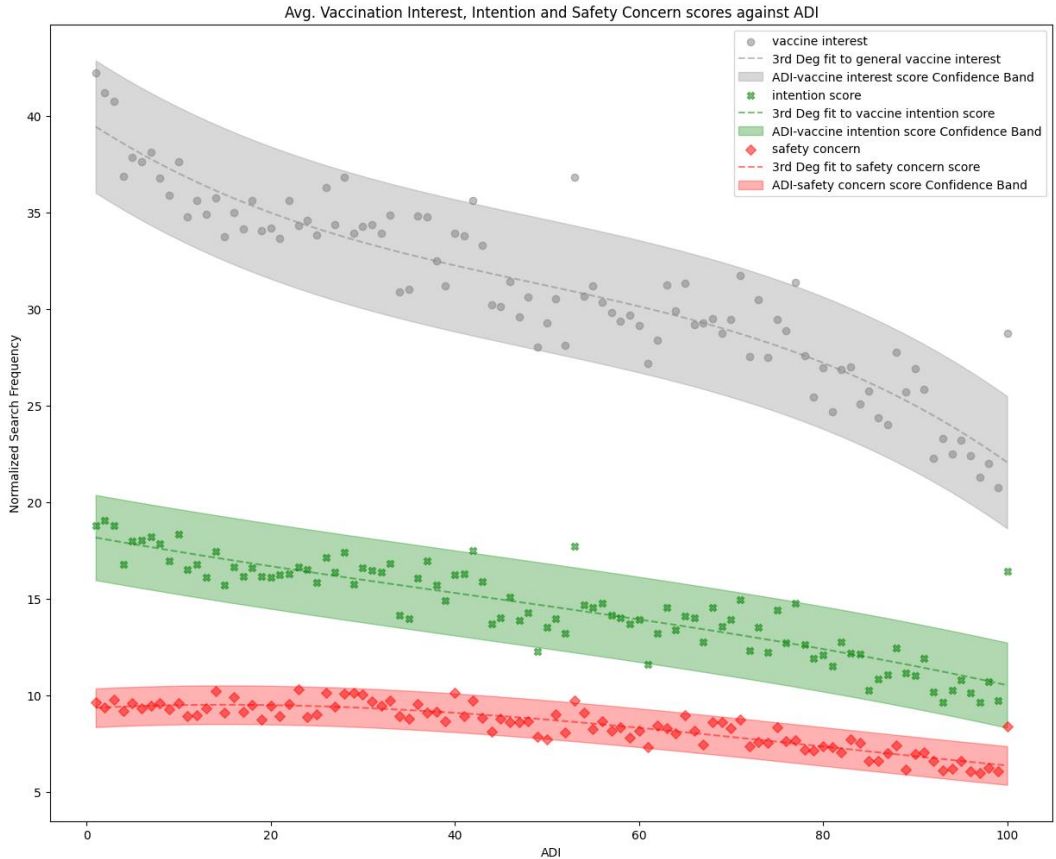

**Figure 3 – Scatter plot of average Covid19 vaccine interest (grey), vaccine intentions (green) and vaccine safety concerns (red) in relative search terms among counties at each ADI percentile, with 3rd order best fit curve and 95% CI.**

All three search trends on Covid19 vaccines exhibit a strong negative linear correlation against ADI (vaccine score: $r_s(100) = -0.920$, intention score: $r_s(100) = -0.872$, safety concern score: $r_s(100) = -0.861$, $p<0.001$). Poorer regions perform fewer relative searches on Covid19 in general, and exhibit less intention and safety concerns regarding receiving the vaccine.

Similarly, polled responses on vaccine willingness showed general positive correlation between hesitancy and deprivation of counties ($r_s(100)=0.767$, $p<0.001$).

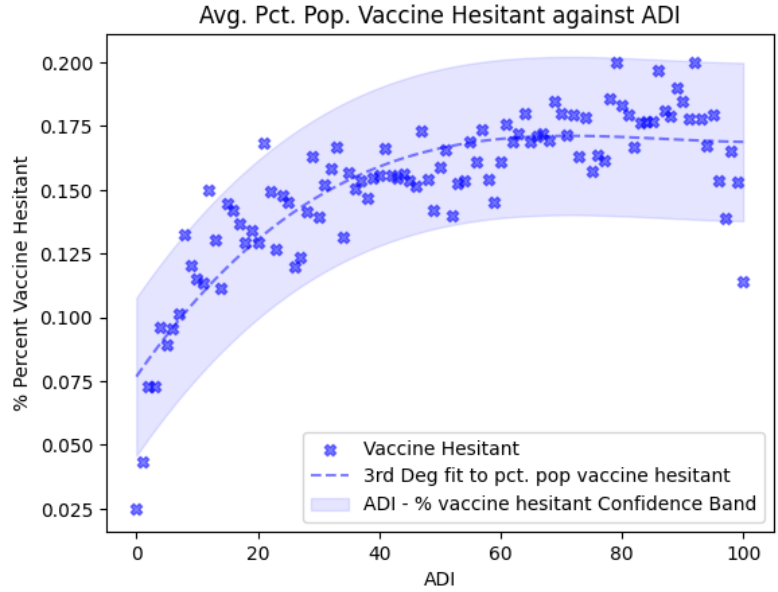

**Figure 4** – Plot of average percentage of population polled as Covid19 vaccine hesitant across counties for each ADI percentile, with best fit curve and 95% CI.

### D. Twitter Sentiment and Mental Health Search Analysis

Unwillingness to comply with societal / official encouragements or discomfort with receiving vaccinations was explored as a possible hypothesis for the above trends. Twitter sentiments relating to Covid19 were used to gauge sentiments – adverse emotions could indicate distrust for Covid19 developments and vaccinations among communities. Scores for negative sentiments of anger, disgust, sadness and fear were normalized against the total number of Tweets in the region.

No positive correlation was found between relative frequency of adverse sentiments on Twitter and higher deprivation. Instead, the wealthiest communities tended to express more adverse emotions through Twitter ($r_s$ = -0.644, -0.627, -0.615, -0.633 for relative frequency of Tweets with anger, disgust, fear and sadness respectively; $p < 0.001$).

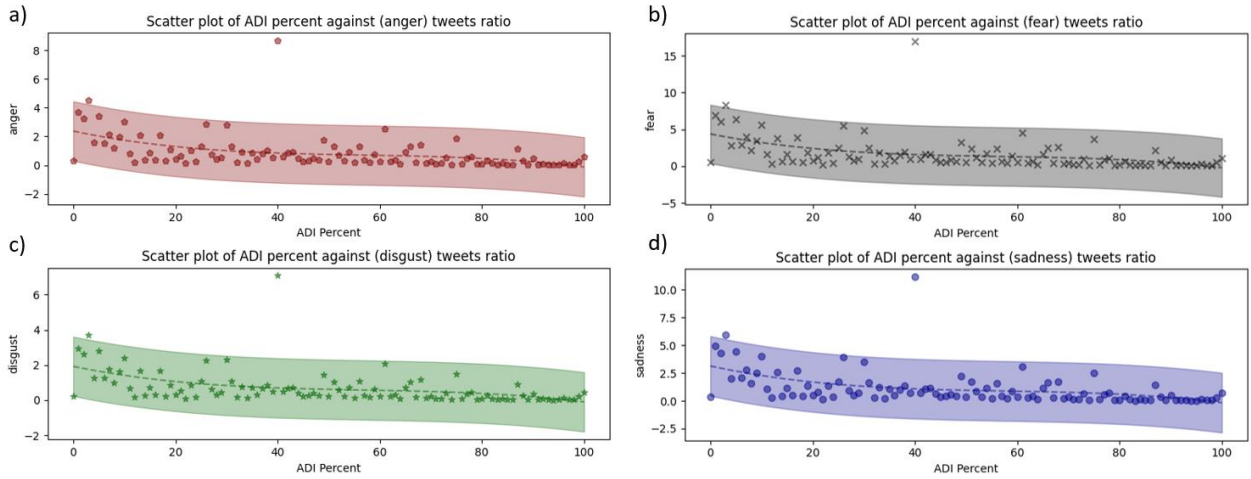

**Figure 5 – Plots of relative sentiments of a): anger, b): fear, c): disgust, d): sadness on Twitter across counties at each ADI percentile, with 95% CI and best fit curve.**

The following plot captures relative search trends for mental health symptoms – panic, depression, insomnia and guilt – against ADI percentile. Mild positive correlations were observed between ADI and panic and insomnia ($r_s$ = 0.491 and 0.575 respectively) while little to no correlation were observed for ADI and depression and guilt ($r_s$ = 0.368 and 0.06 respectively).

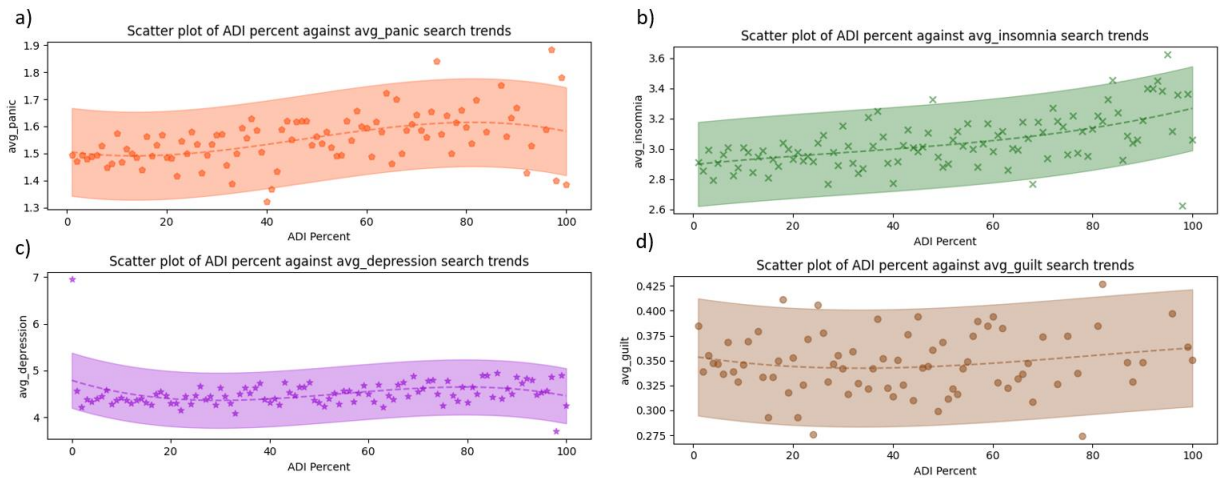

**Figure 6 – Relative frequencies of online searches for a): panic, b): insomnia, c): depression and d): guilt mental health symptoms across counties at each ADI percentile, with 95% CI and best fit curve.**

A possible explanation for higher trends in insomnia, panic searches in higher deprivation regions could be livelihood concerns due to socioeconomic ramifications of the pandemic. The figure below plots average panic and insomnia search trends for counties against food insecurity, discretized by half percentage points of population experiencing food insecurity. Strong positive correlations are noted between food insecurity and panic ($r_s(36)$ = 0.69, p < 0.01) and insomnia ($r_s(36)$ = 0.91, p<0.01).

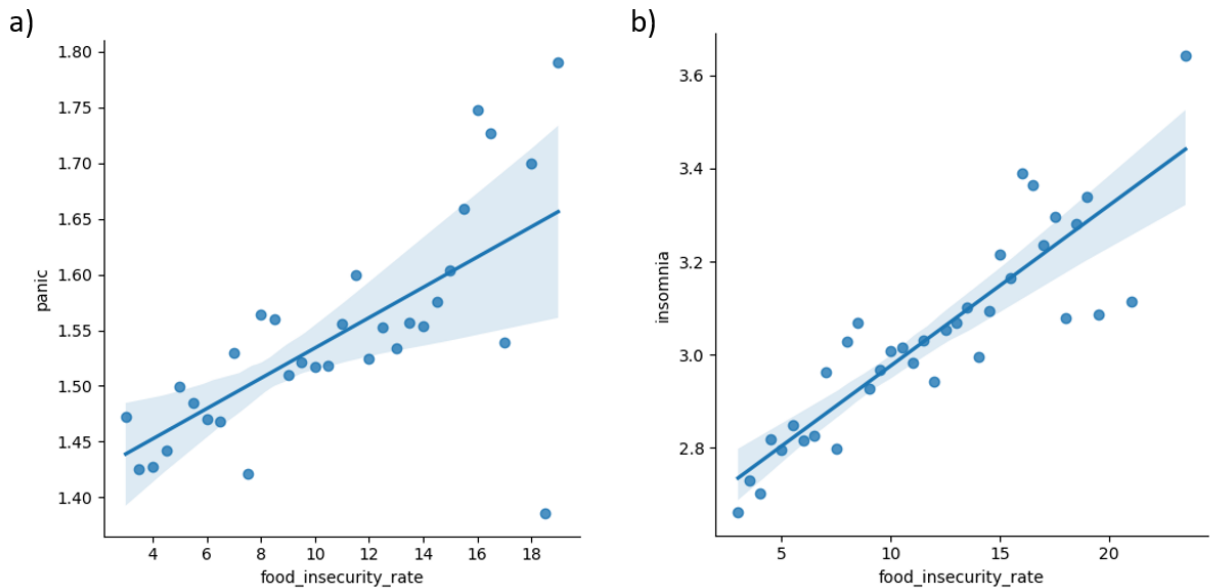

**Figure 7 – Plots of average frequencies of a): panic, b): insomnia mental health search trends in counties at each interval of population percentage experiencing food insecurity.**

E.  Disparities in Polled and Actual Vaccination Rates against County Policies

Data from polls conducted in 2022 can be compared against actual first dose vaccination rates as of 2023 to explore follow-through rates. Unsurprisingly, the polled vaccination rates exceed actual vaccination rates for most counties.

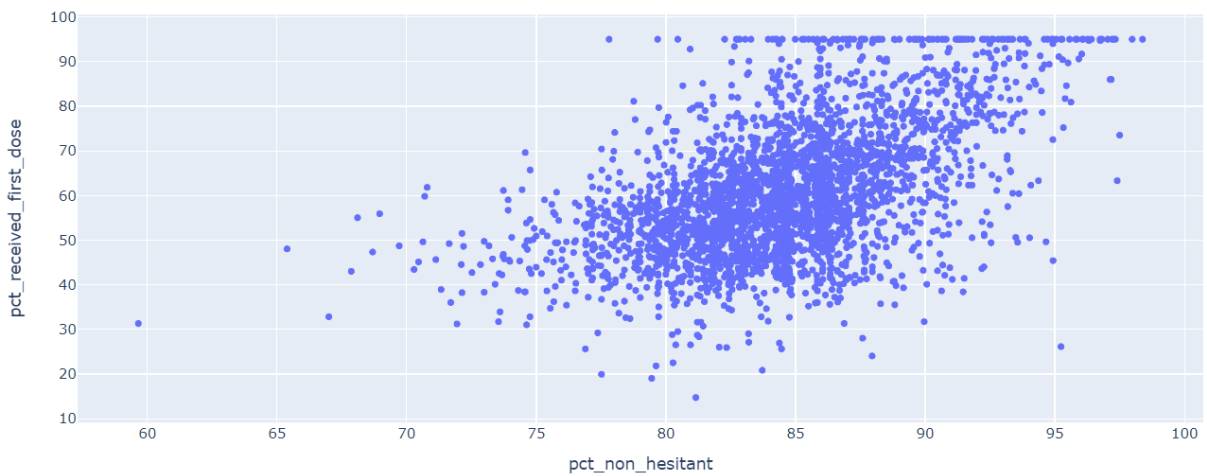

**Figure 8 – Scatter plot of actual first dose vaccination rates against percentage of population polled as willing to receive Covid19 vaccinations. Each datapoint represents a US county.**

Aggregate differences in polled versus actual vaccination rates can be compared for each category of county-level housing assistance, utility assistance and information policy. For each policy, counties with unknown or missing policy values are neglected. The sample sizes for each classification level are as follows: 39 and 59 counties had housing assistance of 0 and 1 respectively, 39 and 46 counties had utility assistance level of 0 and 1 respectively, and 20, 30, 120 counties were categorized as having information policies of 0, 1, 2 respectively. The difference between percentage polled as non-hesitant and actual vaccination rates are shown for counties at each policy classification level. Generally, a smaller difference, indicating lower dropout rate or higher follow-through rate, is desired.

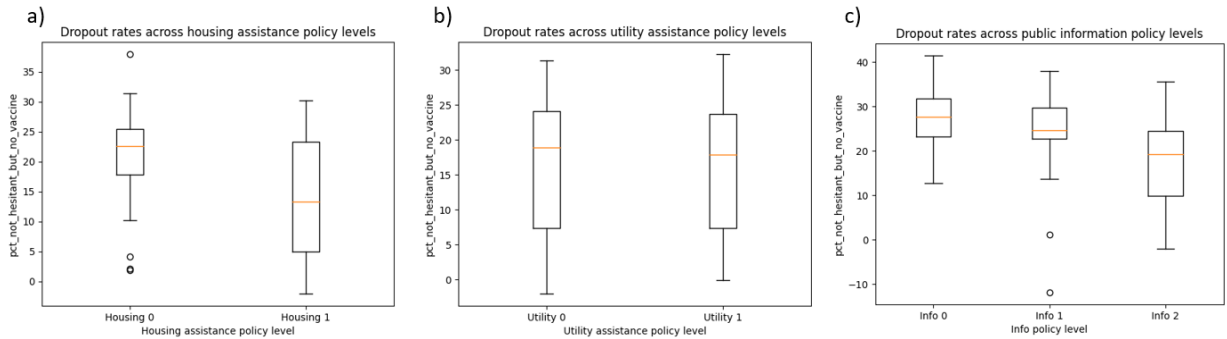

**Figure 9 – Boxplots of dropout rate estimates (difference between percentage polled as not hesitant and percentage who received first dose) at different a): housing assistance, b): utility assistance, c): public information policy levels.**

Counties with stronger housing assistance and information policies were found to have lower disparities between polled willingness and actual vaccination rates. No correlation was observed for utility assistance policies. Stronger public information campaigns (from '0' to '2' categories) was correlated with decreased vaccination dropout rates, from 27.8% to 17.1%. Counties with stronger information policies tended to be wealthier and face lower deprivation, with the mean ADI percentile for counties with '0', '1', and '2' policies at 67.5%, 50.1% and 32.1% respectively.

Introducing housing financial assistance policy similarly correlated with higher follow through rates. Average vaccination dropout rate is lower among counties with housing policies (13.9%) compared to those without (20.6%). Unlike the trend observed in information policies, counties which enacted housing assistance policies had slightly higher deprivation scores (31.5%) than those which did not (28.8%).

## F. Prediction of Covid19 Vaccination Rates

Simple prediction of percentage of first-dose vaccination rates in a county was attempted, using the following features: mental health symptom search trends, ADI, polled hesitancy rates, search interests in Covid19 vaccines, and area and population density of vaccination sites. Random forest regression and linear regression were investigated as supervised learning models. Data was randomly split into training, evaluation and test sets, of sizes 1761, 250 and 495 respectively. Linear regression yielded higher performance on evaluation set and showed a mean absolute error (MAE) of 8.71% and $r^2$ of 0.37. An unsupervised approach using 5-nearest neighbors was implemented on database, with a MAE of 9.04%. In contrast, using the average vaccination rate as prediction yields a MAE of 11.40%. While far from comprehensive, retrospective predictions suggest the possibility of estimating future vaccine acceptance rates using the above features to guide public health policies.

# IV. Conclusions and Limitations

In summary, the following trends were observed:

1. There exists a generally negative correlation between ADI and three measures of vaccination rate (first dose, primary series, and booster dose), although some outliers exist.
2. No clear correlation between travel time, population density of vaccination site and ADI was found.
3. There exists a generally negative correlation between ADI and three measures of vaccination search interest: general interest, intention to vaccinate, and safety concerns.
4. There exists a generally positive correlation between ADI and vaccine hesitancy, although the counties with the highest ADI display lower hesitancy than median ADI counties.
5. No correlation between ADI and the extent of negative sentiments expressed through tweets.
6. There exists a generally positive correlation between searches for specific mental health symptoms (panic, insomnia) and ADI/food insecurity.
7. The exists a generally positive correlation between the presence of specific policies (housing and information), and vaccination follow-through. However, this is based on only a very small sample of counties and may not fully generalize to all US counties.

The above explorations suggests that poorer communities exhibit less interest in and are more hesitant to accept Covid19 vaccines. Although less negative sentiments are outrightly expressed, search trends for metal health symptoms suggest lower mental wellbeing in regions of higher deprivation, possibly due to concerns of effects of the pandemic on livelihoods. Public information campaigns and housing assistance policies correlated with higher rates of follow-through in receiving first doses of Covid19 vaccine, despite the presence of housing policy in particular corresponding to communities with higher deprivation. This supports the use of financial incentives for adoption and adherence to vaccine regimes, especially among communities with higher deprivation [37], [38].

One potential explanation for lower vaccine acceptance among poorer communities is the cognitive burden of poverty [39]–[41]. Poverty and the ensuing mental taxation has been linked to lower quality of modifiable risk factors and higher morbidity and mortality [42]. This could partially explain how protecting basic standards of living through housing assistance policies can increase follow-through rates for Covid19 vaccines. Alternatively, the working poor often experience time poverty [43], [44]; with labor as their main economic asset, these individuals might resort to taking on several jobs to survive, reducing available time and energy to arrange for vaccinations.

Vaccine acceptance among poor communities is a complex issue. Poverty in the US is highly stigmatized [45], and this analysis was conducted through the myopic perspective of highly privileged individuals. Race has been and is a significant factor in inequality [46], and historical and structural inequalities in the US healthcare system has resulted in different races having varying levels of trust in public health policies [47]–[51]. This analysis does not account for racial composition of communities, which is a key consideration for public health policies. Also neglected here is the politicization of vaccines and public health efforts to control Covid19 [52], [53], and the effects of various local policies or grassroot activities [54].

# V. Appendix

Code used in this study can be accessed from: https://github.com/ziiunlai/VaccineAcceptanceDeprivation/.

# Acknowledgments

The authors would like to thank Dr Shiva Shivakumar for introducing us to the world of databases and big data.